\documentclass[10pt,aps,prb]{revtex4}

\usepackage{amsmath}
\usepackage{hyperref}

\begin{document}

\title{Logic and thermodynamics: the heat-engine axiomatics of the second law}
\author{Valery P. Dmitriyev}
\affiliation{Lomonosov University,
P.O. Box 160, Moscow 117574, Russia}
\email{dmitr@cc.nifhi.ac.ru}

\begin{abstract}
We challenge the statement that the principle of Thomson and the
principle of Clausius are equivalent. A logical mistake in the
supposed textbook proof of their equivalency is indicated. On this
account we refine the heat-engine axiomatics. We consider the
energy exchange in the configuration comprised of two heat
reservoirs and one mechanical device and show explicitly the
domains banned by the laws of thermodynamics.
\end{abstract}

\maketitle

In the preface to the textbook \cite{Kubo} on thermodynamics Ryogo
Kubo passed the following observation. "As in contrast to the
atomic theory, thermodynamics does not find a support in our
intuition. This is one of the reasons why students consider
thermodynamics difficult for mastering and can not apply it to
concrete problems." As a confirmation of his words we see
\cite{Dmitriyev} a confusion in the very foundations of the
theory. There is an error in the axiomatics of phenomenological
thermodynamics or, may be, imperfection, that moves over from one
textbook to another for a century (see e.g. Ref.
\onlinecite{Huang, Munster, Goldstein}). This is the well-known
proof by contradiction of that the principle of Thomson is
equivalent to the principle of Clausius and vice versa. This
theorem is mostly decorative and so does not prevent us from the
proper conclusions. However, it may be just the reason of the
difficulties in comprehending the whole structure of
thermodynamics. We use it as a guide to make the formulation of
the second law more coherent and unequivocal. The role of formal
logic in constructing deductive theories is thus emphasized.

We consider the exchange of the energy in the ternary
configuration comprised of the two thermostats $h$ and $c$ and one
mechanical system $m$. The thermomechanical device that implements
this exchange in a cycle is known as the heat engine. The first
law restricts the values $\Delta E$ of the increments of the
energy by the plane
\begin{equation}
\Delta E_h + \Delta E_m + \Delta E_c = 0\, .  \label{1}
\end{equation}
The second law puts further restraints on the vector $(\Delta E_h,
\Delta E_m, \Delta E_c)$. The principle of Thomson forbids the
region in the plane (\ref{1}) where
\begin{equation}
\Delta E_h \leq  0, \,\, \Delta E_m > 0, \,\, \Delta E_c \leq 0
\,. \label{2}
\end{equation}
The principle of Clausius forbids in the plane (\ref{1}) the ray
\begin{equation}
(\Delta E_h, 0, \Delta E_c)\!:\,\, \Delta E_h > 0  \label{3}
\end{equation}
and thus ranks the bodies into cold $c$ and hot $h$. The
principles (\ref{2}) and (\ref{3}) do not specify the whole region
banned by the second law. But they play a crucial role in
constructing this region. As we see, these principles belong to
different domains. So, they can not be equivalent.

Using the both principles we may prove \cite{Huang} the following
lemma:
\begin{equation}
\textrm{if}\, \,\Delta E_m > 0\,\,\, \textrm{then} \,\, \Delta E_c
> 0\,. \label{4}
\end{equation}
Indeed, let $\Delta E_c = 0$. Then we have from Eq.~(\ref{1})
$\Delta E_h < 0$, that contradicts the Thomson's principle
(\ref{2}). Let $\Delta E_c < 0$. The Thomson's principle (\ref{2})
does not forbid the process
\begin{equation}
\Delta E_h >  0,\,\, \Delta E_m > 0,\,\, \Delta E_c < 0 \,.
\label{5}
\end{equation}
It also allows a process
\begin{equation}
\Delta' E_h >  0,\,\, \Delta' E_m < 0,\,\, \Delta' E_c = 0\,.
\label{6}
\end{equation}
Summing up (\ref{5}) and (\ref{6}) for $\Delta' E_m = -\, \Delta
E_m$ we get
\begin{equation}
\Delta E_h + \Delta' E_h
> 0,\,\, \Delta E_m + \Delta' E_m = 0,\,\, \Delta E_c + \Delta' E_c <
0\,. \label{7}
\end{equation}
The process (\ref{7}) contradicts the Clausius' principle
(\ref{3}). That proves lemma (\ref{4}).

To find the region allowed by thermodynamics we accept the
postulate of \textit{the existence of reversible processes} ($r$).
It states that for each $\Delta E_h^{(r)}$ there exists $\Delta
E_c^{(r)}$ such that we can find $\Delta' E_c^{(r)}$ and $\Delta'
E_h^{(r)}$ complying with
\begin{eqnarray}
\Delta E_h^{(r)} + \Delta' E_h^{(r)} &=& 0\label{8}\\
\Delta E_c^{(r)} + \Delta' E_c^{(r)} &=& 0\,. \label{9}
\end{eqnarray}
The reverse process is unique that is easily proved using the
Thomson's principle (\ref{2}).

Proceeding from (\ref{4}) and (\ref{8}), (\ref{9})  we may prove
the Carno's theorem. It says that the efficiency of a reversible
heat engine is greater or equal to the efficiency of any other
heat engine. Below we give a more accurate proof of the Carno's
theorem than it is usually done \cite{Kubo}.

Consider the energy exchange in the ternary configuration
according to Eq.~(\ref{1}). We want to compare (\ref{1}) with the
reversible process
\begin{equation}
\Delta E_h^{(r)} + \Delta E_m^{(r)} + \Delta E_c^{(r)} = 0
\label{10}
\end{equation}
under the condition
\begin{equation}
\Delta E_h = \Delta E_h^{(r)}\,. \label{11}
\end{equation}
To this end, we will carry out firstly (\ref{1}) and then the
process
\begin{equation}
\Delta' E_h^{(r)} + \Delta' E_m^{(r)} + \Delta' E_c^{(r)} = 0
\label{12}
\end{equation}
that is reverse to (\ref{10}). Summing up Eq.~(\ref{1}) and
Eq.~(\ref{12}) and using in it (\ref{11}) and (\ref{8}) we get
\begin{equation}
\Delta E_m + \Delta' E_m^{(r)} + \Delta E_c + \Delta' E_c^{(r)} =
0\,. \label{13}
\end{equation}
By virtue of the Thomson's principle (\ref{2}) we have from
Eq.~(\ref{13})
\begin{equation}
\Delta E_c + \Delta' E_c^{(r)} \geq 0\,. \label{14}
\end{equation}
Excluding $\Delta' E_c^{(r)}$ from Eq.~(\ref{14}) and (\ref{9})
gives
\begin{equation}
\Delta E_c - \Delta E_c^{(r)} \geq 0\,. \label{15}
\end{equation}
Let $\Delta E_m^{(r)} > 0$. Then by the lemma (\ref{4}) $\Delta
E_c^{(r)} > 0$. Inequality (\ref{15}) says that with the condition
(\ref{11}) the work made by the reversible heat engine is greater
than the work done by the irreversible heat engine. Dividing
(\ref{15}) by $\Delta E_c^{(r)}$ and using in it tautologically
(\ref{11}) we get
\begin{equation}
-\,\frac{\Delta E_h}{\Delta E_h^{(r)}} + \frac{\Delta E_c}{\Delta
E_c^{(r)}}\, \geq 0\,. \label{16}
\end{equation}
Let $\Delta E_m^{(r)} < 0$. Then by the lemma (\ref{4}) and
(\ref{9}) we have for reversible processes $\Delta E_c^{(r)} < 0$.
In this case inequality (\ref{16}) should be replaced by
\begin{equation}
\,\frac{\Delta E_h}{\Delta E_h^{(r)}} - \frac{\Delta E_c}{\Delta
E_c^{(r)}}\, \geq 0\,. \label{17}
\end{equation}
At last, we may unite (\ref{16}) and (\ref{17}) into the
inequality
\begin{equation}
\,\frac{\Delta E_h}{|\Delta E_h^{(r)}|} + \frac{\Delta
E_c}{|\Delta E_c^{(r)}|}\, \geq 0\,. \label{18}
\end{equation}

Combine $n$ irreversible and $n^{(r)}$ reversible heat engines
such that
\begin{equation}
\frac{\Delta E_h}{\Delta E_h^{(r)}} =
\frac{\,\,\,\,\,n^{(r)}}{n}\, . \label{19}
\end{equation}
Then we may deduce \cite{Huang} (\ref{18}) for a general case that
does not obey the restriction (\ref{11}). As you see, the proof of
the Carno's theorem makes use of the Thomson's principle
(\ref{2}). It can also be proved using the Clausius' principle
(\ref{3}).

Let us take instead of (\ref{1}) a reversible process $(r_0)$.
Then, reversing the processes,  we may turn \cite{Kubo} (\ref{18})
into equality. This equality can be written as
\begin{equation}
\frac{|\Delta E_h^{(0)}|}{|\Delta E_h^{(r)}|} = \frac{|\Delta
E_c^{(0)}|}{|\Delta E_c^{(r)}|}\,. \label{20}
\end{equation}
Now we can define \cite{Kubo, Huang} an absolute temperature scale
by the relations
\begin{equation}
|\Delta E_h^{(0)}|= \kappa\, \theta_h\,, \quad |\Delta E_c^{(0)}|
= \kappa\, \theta_c \label{21}
\end{equation}
where $\kappa > 0$ is a constant. Substituting (\ref{20}) and
(\ref{21}) into (\ref{18}) gives
\begin{equation}
\frac{\Delta E_h}{\theta_h} + \frac{\Delta E_c}{\theta_c}\, \geq
0\,. \label{22}
\end{equation}
Taken together with Eq.~(\ref{1}), inequality (\ref{22}) specifies
 the domain of the processes allowed by thermodynamics. This is a
half-plane in the energy space $(\Delta E_h, \Delta E_m, \Delta
E_c)$.

As we see from the above, the principle of Thomson and the
principle of Clausius are not equivalent. However, there is a
popular textbook proof \cite{Huang, Munster, Goldstein} of their
equivalence. In order to show explicitly the error in the supposed
proof we must give both the verbal and logical formulation of
these principles. The Thomson's principle states: One can not take
heat from a body and totally convert it to work. The Clausius'
principles states: "Heat cannot, of itself, pass from a colder to
a hotter body." The equivalence is proved by contradiction (i.e.
from the contrary). The two theorems are needed.

A: If the Thomson's principle is wrong then the Clausius'
principle is wrong as well.

B: If the Clausius' principle is wrong then the Thomson's
principle is wrong as well.

\noindent It is sufficient to analyze the supposed proof of the
first theorem. It runs as follows.

1. Let the Thomson's principle be wrong.

2. Then we may take a heat from the cold body and convert it
wholly to a work.

3. Next we may convert this work into the internal energy  of the
hot body.

4. The pure result of these operations is that heat was
transferred from the cold body to the hot body.

\noindent That contradicts the Clausius' principle and hence the
statement A is valid.

Here the error is in the step 3. To explicate it we formulate the
Thomson's principle in logical terms. Let us define on the line
$\Delta E + \Delta E_m =0$ the Boolean function that describes the
energy exchange between a termostat and a mechanical system:
\begin{eqnarray}
I(\Delta E_m) &=& true \,\,\, \,\,\,\, \textrm{for} \,\,\,\ \Delta E_m \leq  0 \label{23}\\
I(\Delta E_m) &=& false \,\,\,\, \textrm{for} \,\,\,\ \Delta E_m
> 0\,. \label{24}
\end{eqnarray}
This function just expresses the Thomson's principle. Negating it
we get
\begin{eqnarray}
^\neg I(\Delta E_m) &=& false \,\,\,\, \textrm{for} \,\,\,\ \Delta E_m \leq 0 \label{25}\\
^\neg I(\Delta E_m) &=& true \,\,\, \,\,\,\, \textrm{for} \,\,\,\
\Delta E_m
> 0\, . \label{26}
\end{eqnarray}
This allows $\Delta E_m > 0$ and forbids $\Delta E_m \leq 0$. We
made use (\ref{26}) in step 2 and violated (\ref{25}) in step 3.
The origin of this mistake is in that the verbal formulation of
the Thomson's principle says explicitly (\ref{24}) and does not
speak out (\ref{23}). However, the latter is implicit as in an
ellipsis or enthymeme. The literal understanding of the words gave
rise to the error in the logical inference. This illustrates the
significance of formal logic in constructing deductive theories.


\begin{thebibliography}{99}

\bibitem{Kubo} Ryogo Kubo et al., \textit{Thermodynamics. An advanced course with problems and
solutions}, North Holland Publishing Company, Amsterdam 1968.
\bibitem{Dmitriyev} V.P.Dmitriyev, ``The independence and mutual complementarity of the principles by Thomson and
by Clausius'', J. Phys. Chem. \textbf{59} (1), 41 (1985), In
Russian.
\bibitem{Huang} Kerson Huang, \textit{Statistical mechanics}, John Wiley
and Sons, New Work - London 1963, Part A.
\bibitem{Munster} A.Munster, \textit{Classical thermodynamics}, John Wiley and
Sons,
1970.

\bibitem{Goldstein} M.Goldstein, ``A preface to the Carno cycle'', J. Chem. Ed. \textbf{57} (2), 114 (1980).


\end{thebibliography}
\end{document}